\documentclass{article}

\makeatletter

\@addtoreset{equation}{section}
\makeatother

\usepackage{amsmath,amssymb}
\usepackage{cite}
\usepackage{graphicx}
\usepackage[vcentermath]{youngtab}
\Yboxdim{4pt}

\newcommand {\beq}{\begin{eqnarray}}
\newcommand {\eeq}{\end{eqnarray}}

\newcommand{\bC}{\ensuremath{\mathbb{C}}}

\newcommand{\bH}{\ensuremath{\mathbb{H}}}

\newcommand{\bO}{\ensuremath{\mathbb{O}}}
\newcommand{\bP}{\ensuremath{\mathbb{P}}}

\newcommand{\bR}{\ensuremath{\mathbb{R}}}

\newcommand{\scA}{\ensuremath{\mathcal{A}}}

\newcommand{\scC}{\ensuremath{\mathcal{C}}}

\newcommand{\scN}{\ensuremath{\mathcal{N}}}

\newcommand{\scW}{\ensuremath{\mathcal{W}}}

\newcommand{\frakg}{\ensuremath{\mathfrak{g}}}

\begin{document}
\baselineskip 0.7cm

\begin{titlepage}

\setcounter{page}{0}

\renewcommand{\thefootnote}{\fnsymbol{footnote}}

\begin{flushright}
UT-08-28\\
IPMU-08-0063\\
September, 2008
\end{flushright}

\vskip 1.35cm

\begin{center}
{\Large \bf
Octonions, $G_2$ and generalized Lie 3-algebras
}

\vskip 1.2cm 

{\normalsize
Masahito Yamazaki
}

\vskip 0.8cm

{ \it
Department of Physics, University of Tokyo, Hongo 7-3-1, Tokyo, 113-0033, Japan
\\
yamazaki(at)hep-th.phys.s.u-tokyo.ac.jp
}

\end{center}

\vspace{12mm}

\centerline{{\bf Abstract}}
We construct an explicit example of a generalized Lie 3-algebra from the octonions. In combination with the result of \cite{CS}, this gives rise to a three-dimensional $\scN=2$ Chern-Simons-matter theory with exceptional gauge group $G_2$ and with global symmetry $SU(4)\times U(1)$. This gives a possible candidate for the theory on multiple M2-branes with $G_2$ gauge symmetry.
\end{titlepage}

\section{Introduction}
Recently, tremendous research activities have been triggered by the seminal works of Bagger, Lambert and Gustavsson (BLG) \cite{BLG}, who proposed a Chern-Simons type
Lagrangian describing the world-volume theory of multiple M2-branes. 
The original BLG theory uses algebraic structure known as Lie 3-algebras (and nonassociative algebras), and they
used a single example of a Lie 3-algebra known as $\scA_4$ . Unfortunately it was later proven \cite{no-go}
 that $\scA_4$ is essentially the unique possibility if we impose positivity of the metric. Given this no-go theorem, one possible way to proceed is to lift the condition that the metric is positive definite. Theories
with Lorentzian metrics are proposed in \cite{Lorentzian}, and they can realize arbitrary gauge groups.
However, the problem of this approach is that the unitarity of these theories is often not clear, and one method of making the theory unitary \cite{unitary} leads us back to the original $\scN=8$ super Yang-Mills theory \cite{D2toD2}.
    
In another line of development, Aharony, Bergman, Jafferis and Maldacena
(ABJM) \cite{ABJM} proposed three-dimensional Chern-Simons matter theories with $\scN = 6$ supersymmetry with gauge groups $U (N)\times U(N)$ and $SU(N) \times SU(N)$.\footnote{In this paper, we do not bother about the differences of the Lie group $G$ and its associated Lie algebra $\frakg$.} Although original ABJM approach does not use Lie 3-algebras, 
 it was later shown
in \cite{BL4} that ABJM theory can be understood as a special class of models which arises by considering generalized algebraic structures (which is called Hermitian 3-algebras in \cite{recent}). This in particular abandons the
antisymmetricity of the structure constants. 

More recently, another interesting paper \cite{CS}
appeared, which proposes a class of new theories using yet another generalization of Lie 3-algebras (called generalized Lie 3-algebras) whose structure constants
are not antisymmetric.
Based on generalized Lie 3-algebras, they have constructed three-dimensional Chern-Simons-matter theories with $\scN=2$ supersymmetry and with manifest $SU (4) \times  U (1)$ global symmetry \footnote{The claim of \cite{CS} is that we have $SU(4)\times U(1)$ R-symmetry, and thus have $\scN=6$ supersymmetry. As we will discuss later, this is actually not the case and this $SU(4)\times U(1)$ symmetry is in general just a global symmetry.} for each generalized Lie 3-algebra. As examples, they give a class
of theories called $\scC_{2d}$, but clearly it is an interesting problem to see whether we have more
examples of theories of this type. From our experience of metric Lie 3-algebras (for which we have the no-go theorem \cite{no-go}), we expect that this is a highly non-trivial problem.

    In this paper, we give an explicit example of a generalized 3-Lie algebra by using the octonions. When this algebra is combined with the result of \cite{CS}, we have a possible candidate
for the theory on the worldvolume of M2-branes with exceptional gauge group $G_2$. The appearance of exceptional gauge group is particularly interesting, since we do not have such a theory in three-dimensional theories with higher supersymmetries. For example, in ABJM-type theory with $\scN=6$ supersymmetry, the result of \cite{ST} shows that the gauge group is necessarily $SU (n) \times U (1), Sp(n) \times U (1), SU (n) \times  SU (n)$ and $
SU (n) \times  SU (m) \times  U(1)$ with possibly additional $U (1)$'s. In theories with $\scN=5$ supersymmetry, the gauge group is either $SO(m)\times Sp(n)$ \cite{Hosomichi}, $SO(7)\times Sp(1), G_2\times Sp(1)$ or $SO(4)\times Sp(1)$\cite{Bergshoeff}. In particular, when we want to have exceptional gauge group $G_2$ with $\scN=5$ supersymmetry, that $G_2$ is always accompanied by an extra $Sp(1)$, whereas we have gauge group purely $G_2$ in the theory we consider in this paper.

The organization of the paper is as follows. In section \ref{sec.review}, we briefly review generalized Lie 3-algebras and the work of \cite{CS}. In section \ref{sec.octonion}, we give
an explicit example of a generalized Lie 3-algebra from the octonions. 
In section \ref{sec.M2}, we discuss
the possible physical implication of this result from the viewpoint of the worldvolume
theory on M2-branes. 
Finally, in section \ref{sec.conc} we conclude this paper with summary and discussions. Appendix \ref{sec.appendix} is devoted to some discussions of alternative algebras.

\section{Generalized Lie 3-algebras} \label{sec.review}
Let us first give the definition of a generalized metric Lie 3-algebra following \cite{CS} \footnote{In \cite{CS}, this is called a generalized metric Lie 3-algebra with pairings. Our terminology here is just for simplicity.}.

\paragraph{Definition.} A {\em generalized metric 3-Lie algebra} consists of an  algebra $\scA$ with a ternary map $[\cdot,\cdot,\cdot]:\scA^3\rightarrow \scA$ and a symmetric, bilinear, positive definite pairing $(\cdot,\cdot):\scA^2\rightarrow \bR$ satisfying the following properties:

1) {\em fundamental identity:}
\begin{equation}\label{fundamentalidentity}
 {}[x,y,[a,b,c]]\ =\ [[x,y,a],b,c]+[a,[x,y,b],c]+[a,b,[x,y,c]],
\end{equation}

2) invariance of the pairing or {\em metric compatibility condition:}
\begin{equation}\label{metriccompatibilitygen}
 {}([x,y,a],b)+(a,[x,y,b])\ =\ 0,
\end{equation}

3) the additional symmetry property:
\begin{equation}
 ([x,y,a],b)\ =\ ([a,b,x],y),
\end{equation}

\noindent for all $x,y,a,b,c\in\scA$.

We can rewrite these conditions in terms of structure constants. Structure constants ${f^{abc}}_d$ are introduced just as in the case of Lie 3-algebras:
\begin{equation}
 [e^a,e^b,e^c]\ =\ f^{abc}{}_de^d. 
\end{equation}
where $e^a$'s are basis of $\scA$ and $h^{ab}=(e^a, e^b)$.

The conditions we imposed above on generalized metric Lie 3-algebras can be reformulated using the structure constants. The fundamental identity reads as
\begin{equation}
 f^{efg}{}_df^{abc}{}_g\ =\ f^{efa}{}_gf^{gbc}{}_d+f^{efb}{}_gf^{agc}{}_d+f^{efc}{}_gf^{abg}{}_d~,
\end{equation}
and the remaining conditions are captured by the symmetry properties
\footnote{The structure constants having these symmetry properties are mentioned briefly in the Appendix C.5 of \cite{Gustavsson}, although the example of structure constants we discuss below is certainly new.}
\begin{equation}
 f^{abcd} = -f^{bacd} = -f^{abdc} =f^{cdab}.
\end{equation}\\


Given a generalized Lie 3-algebra, we can write down the Lagrangian of the theory, using four sets $\Phi^i, \bar{\Phi}^i\ (i=1,\ldots,4)$ of chiral superfields and a vector superfield $V$. Here we do not write down the explicit form of the action; see \cite{CS} for details. Since the action is written in $\mathcal{N}=2$ superfield formalism, it is clear that we have $\scN=2$ supersymmetry. Global symmetry $SU(4)\times U(1)$ is also manifest from the form of the action. 



At this point some readers might wonder whether $\scN=2$ supersymmetry in three dimensions is enough to determine the form of the Lagrangian completely. In fact, the answer is definitely no. The possible ambiguity resides in the form of the superpotential $\scW(\Phi)$, which should be a polynomial in $\Phi$ and should be constructed from the triple bracket and the metric. Possible forms of $\scW$ include:
\begin{equation}
 \scW_\alpha(\Phi)\ =\ \alpha\epsilon_{ijkl}([\Phi^i,\Phi^j,\Phi^k],\Phi^l)\quad \textrm{and}\quad
 \scW_\beta(\Phi)\ =\ \beta([\Phi^i,\Phi^j,\Phi^i],\Phi^j)~.\label{eq.W}
\end{equation}
The first term is similar to the potential term of the BLG theory, whereas the second term is absent in the ordinary (non-generalized) Lie 3-algebras. In this sense, the theories of \cite{CS} are different from BLG(-type) theories constructed from ordinary Lie 3-algebras. 

%

\section{An example of a generalized Lie 3-algebra from octonions} \label{sec.octonion}
In the previous section, we explained general formalism applicable to arbitrary generalized metric Lie 3-algebras. However, we still have to find explicit examples of generalized Lie 3-algebras to complete the story. In \cite{CS}, the only example discussed in \cite{CS} is the algebra $\scC_{2d}$, and we are definitely in need of more explicit examples of generalized Lie 3-algebras. 
In this section, we thus give an explicit construction of a generalized Lie 3-algebra using the octonions, which is one of the most famous non-associative algebras in the literature\footnote{See \cite{oldoct} for examples of old discussions of octonions in the context of membranes and \cite{Mauri} for more recent discussions in the context of BLG theories. However, these works use the so-called ``octonionic structure constants'' $\phi_{ijk}$ defined by
\beq
e_i e_j=-\delta_{ij}+\phi_{ijk}e_k,
\eeq
or its dual $\phi_{ijkl}$ in seven dimensions
\beq
\phi_{ijkl}=\epsilon_{ijklmno}\phi_{mno},
\eeq
which is different from our structure constants $f^{abcd}$. If we use the notation shown below \eqref{eq:decomp}, then the antisymmetric part $\Phi_F$ of $f^{abcd}$ corresponds to $\phi_{ijkl}$.}.

Octonions $\bO$ are one of the four normed division algebras (the other three are $\bR, \bC$ and $\bH$). It is an important example of alternative algebra (see Appendix for definition of alternative algebras), which is a special class of non-associative algebra. It is spanned by 8 basis $e_a (a=0,1,\ldots, 7)$, whose multiplication table is given in Table \ref{tbl.oct}. Here we are taking $e_0$ to be an identity. 

\begin{table}[htbp]
\caption{multiplication table of octonions. }
\begin{center}
\begin{tabular}{c|cccccccc}
 &$e_0$ &  $e_1$ & $e_2$  & $e_3$ & $e_4$ & $e_5$ & $e_6$ & $e_7$ \\
\hline
$e_0$ & $e_0$ &  $e_1$ & $e_2$  & $e_3$ & $e_4$ & $e_5$ & $e_6$ & $e_7$ \\

$e_1$& $e_1$ & $-e_0$ & $e_4$  & $e_7$ & $-e_2$ & $e_6$ & $-e_5$ & $-e_3$ \\

$e_2$ &$e_2$ & $-e_4$ & $-e_0$  & $e_5$ & $e_1$ & $-e_3$ & $e_7$ & $-e_6$ \\

$e_3$ & $e_3$ & $-e_7$ & $-e_5$  & $-e_0$ & $e_6$ & $e_2$ & $-e_4$ & $e_1$ \\

$e_4$ & $e_4$ &$e_2$ & $-e_1$  & $-e_6$ & $-e_0$ & $e_7$ & $e_3$ & $-e_5$ \\

$e_5$ & $e_5$ &$-e_6$ & $e_3$  & $-e_2$ & $-e_7$ & $-e_0$ & $e_1$ & $e_4$ \\

$e_6$ & $e_6$ &$e_5$ & $-e_7$  & $e_4$ & $-e_3$ & $-e_1$ & $-e_0$ & $e_2$ \\

$e_7$ & $e_7$ &$e_3$ & $e_6$  & $-e_1$ & $e_5$ & $-e_4$ & $-e_2$ & $-e_0$ 
\end{tabular}
\end{center}
\label{tbl.oct}
\end{table}

It is easy to check that this algebra is nonassociative; for example, $(e_1 e_2)e_3= e_4 e_3=-e_6$ but $e_1(e_2 e_3)=e_1 e_5=e_6$.\\



For $x \in \bO$, define $R_x, L_x: \bO\to \bO$ by
\beq
R_x(y)=xy, \quad L_x(y)=yx
\eeq
for $y\in \bO$.

Define $D_{x,y}$ by
\begin{equation}
D_{x,y}=[L_x,L_y] +[L_x,R_y]+[R_x,R_y],
\end{equation}
where bracket $[-,-]$ is the commutator as usual. More explicitly, 
\beq
D_{x,y}(z)=x(yz)-y(xz)+(zy)x-(zx)y+(xz)y-x(zy).
\eeq
From this definition, we have
\beq
D_{x,y}=-D_{y,x}. \label{eq.interchange}
\eeq


Now the important fact is that this $D_{x,y}$ is actually a derivation \cite{schafer}. In order words, 
\beq
D_{x,y}(zw)=(D_{x,y}z)w+z D_{x,y}(w) \label{eq.derivation}
\eeq
for all $x,y, z,w\in \scA$. In fact, it is known \cite{schafer} that any derivation $D$ can be written as the sum of $D_{x,y}$:
\beq
D=\sum_{x,y} c_{x,y} D_{x,y}.
\eeq
for some constants $c_{x,y}$.

Let us now define our 3-bracket by \footnote{This definition of 3-bracket appeared already in the classic paper by Nambu \cite{Nambu} in 1973.}
\beq
[x,y,z]:=D_{x,y}(z).\label{eq.bracket}
\eeq
By explicit computations, it is again easy to check that this 3-bracket is not totally antisymmetric, although it is antisymmetric with respect to the interchange of $x$ and $y$ due to \eqref{eq.interchange}. 

Having defined our 3-bracket, we now have to verify that our 3-bracket satisfies the fundamental identity. The key to prove this is \eqref{eq.derivation}. In fact, $[x,y,[a,b,c]]=D_{x,y}(D_{a,b}(c))$, and if we use the explicit expression, $D_{a,b}(c)$ is a sum of products of $a$, $b$ and $c$. Now $D_{x,y}$ acts as a derivation for each of these terms, and thus on the whole $D_{a,b}(c)$. This proves the fundamental identity.\\

Of course, we have to define a metric on $\bO$ before verifying other axioms.
Since $\bO$ is a normed algebra, we have a natural metric. For an element $x=x^0 e^0+\ldots x^7 e^7 \in \bO$, define its conjugate $x^*$ by $x^*=x^0 e^0-x^1 e^1-x^2 e^2-\ldots -x^7 e^7$. Then the metric on $\bO$ is defined by
\beq
(x,y)=\textrm{Re}(x^* y).\label{eq.metric}
\eeq
where $\textrm{Re}$ is defined by taking the $e_0$ component:
\beq
\textrm{Re}\left(\sum_{a=0}^7 x^a e^a\right)=x^0.
\eeq
In particular, for basis $e^i, e^j$, we have
\beq
(e^i,e^j)=\delta_{ij}.
\eeq
and in general
\beq
(x,y)=\sum_{a=0}^7 x^a y^a
\eeq
This metric is clearly symmetric in $x$ and $y$. Moreover, since the metric is diagonal in our basis, we will hereafter not worry about the differences of upper and lower indices.

Now the metric defined above satisfies
\beq
(xy,z)=(x,yz)
\label{eq.change}
\eeq
for all $x,y,z \in \bO$. To prove this, it suffices to show that 
\beq
(e^i e^j, e^k)=(e^i, e^j e^k). \label{eq.move}
\eeq
We only need to verify this when $(e^i e^j, e^k)\ne 0$, namely $e^i e^j=\pm e^k$. From the multiplication table (Table \ref{tbl.oct}) we learn that $e^j e^k=\pm e^i$, thus we have proven \eqref{eq.move}.


From \eqref{eq.change}, we can directly that
\beq
([x,y,z],w)=-([x,y,w],z).
\eeq
In fact,
\begin{equation*}
\begin{split}
([x,y,z],w)&=(x(yz)-y(xz)+(zy)x-(zx)y+(xz)y-x(zy),w) \\
&=(wx,yz)-(wy,xz)+(zy,xw)-(zx,yw)+(xz,yw)-(wx,zy) \\
&=[(wx,yz)-(zx,yw)]+[(zy,xw)-(wy,xz)]+[(xz,yw)-(xw,yz)]
\end{split}
\end{equation*}
and it is clear from the final expression that the result is antisymmetric in $z,w$. Note that in the final line, we have used the identity
\beq
(xy,zw)=(yx,wz),
\eeq
which can again be verified by similar considerations as in \eqref{eq.change}.

It is another straightforward exercise of octonions to verity that the remaining axiom of generalized Lie 3-algebras:
\beq
([x,y,a],b)=([a,b,x],y).
\eeq

Summarizing, if we define 3-bracket by \eqref{eq.bracket} and metric by \eqref{eq.metric}, then all the conditions of generalized metric 3-Lie algebra is satisfied. In the next section we will see that this algebra gives an interesting theory when applied to the formalism of \cite{CS}.\\

Before finishing this section, it is probably instructive to comment on the relation of our 3-bracket with previous approaches. Since the 3-bracket $[x,y,z]$ is antisymmetric in first two indices, the 3-bracket defines a linear map $\Phi:\wedge^2 \scA\otimes \scA \to \scA$. By decomposing $\wedge^2 \scA \otimes \scA$ into 
\beq
\wedge^2 \scA \otimes \scA=\wedge^3 \scA \oplus \scA^{\yng(2,1)}, \label{eq:decomp}
\eeq
$\Phi$ is broken down into two components $\Phi_F$ and $\Phi_L$ \cite{recent}:
\begin{enumerate}
\item $\Phi_F : \Lambda^3V \to V$, which is totally skewsymmetric.
\item $\Phi_L : V^{\yng(2,1)} \to V$, which is such that
\begin{equation}
    \Phi_L(x,y,z) + \Phi_L(z,x,y) + \Phi_L(y,z,x) = 0~.\label{eq.Lietriple}
\end{equation}
\end{enumerate}
When $\Phi=\Phi_F$ (i.e. $\Phi_L=0$), $[x,y,z]$ is totally antisymmetric and we have an ordinary Lie 3-algebra as discussed in \cite{BLG}. When $\Phi=\Phi_L$ (i.e. $\Phi_F=0$), the algebraic structure satisfying \eqref{eq.Lietriple} is called Lie triple systems (see \cite{Nilsson} for a recent discussion). For our 3-bracket $[x,y,z]$, $\Phi_F=3 \langle x,y,z \rangle \ne 0$ (here we used \eqref{eq.alt} in Appendix), where $\langle x,y,z \rangle $ is the so-called associator which is defined by
\beq
\langle x,y,z \rangle =x(yz)-(xy)z.
\eeq
Now recall that the Bagger-Lambert paper \cite{BLG} constructs 3-brackets from the nonassociative algebras by
\beq
[x,y,z]_{\mathrm{BL}}:=\langle x,y,z \rangle \pm (\mbox{perm.}).
\eeq
Since associator of octonions is antisymmetric with respect to its three arguments (see Appendix), we learn that our $\Phi_F(x,y,z)$ and $[x,y,z]_{\mathrm{BL}}$ coincide. However, we have another piece $\Phi_L(x,y,z)=[x,y,z]-3 \langle x,y,z \rangle \ne 0$, and our 3-bracket is a mixture of the above two. The fundamental identity is not satisfied by either $\Phi_F$ and $\Phi_L$, and only by their sum $\Phi$.  This explains the similarities and differences of our approach and the approach taken by \cite{BLG}.

\section{M2-branes with gauge group $G_2$?} \label{sec.M2}

Having obtained an example of generalized 3-Lie algebra, we turn to the physical implication of this result. 

Let us study the gauge symmetry of our theory. For that purpose, we need the gauge transformation of gauge fields:
\beq
\delta \tilde{A}_{\mu}{}^a{}_b=\partial_{\mu}\tilde{\lambda}^a{}_{b} +\tilde{A}_{\mu}{}^a{}_c \tilde{\lambda}^c{}_b-\tilde{\lambda}^a{}_c \tilde{A}_{\mu}{}^c{}_b,
\eeq
where $\tilde{A}_{\mu}{}^a{}_b=f^{cda}{}_{b}A_{\mu}{}_{cd}$ and $\tilde{\lambda}^a{}_b=f^{cda}{}_b \lambda_{cd}$.
This is the usual gauge transformation with parameter $\tilde{\lambda}^b{}_a$. 
Just as in BLG theory, structure constants are antisymmetric when $a$ and $b$ are exchanged, and thus the gauge group is the subgroup of $SO(\bO)\simeq SO(8)$. But we can indeed say more than that.  

Of course, it is easy to notice that the gauge group is a subgroup of $SO(7)$. This is because our 3-bracket is zero whenever we have an identity $e_0$ in one of its arguments:
\beq
[x,y,e_0]=[x,e_0,y]=[e_0,x,y]=0,
\eeq
or in terms of structure constants,
\beq
f^{0abc}=f^{a0bc}=f^{ab0c}=f^{abc0}=0.\label{eq:decouple}
\eeq
Does this mean that the gauge group is $SO(7)$? The answer is no; explicit computations by Mathematica tells us that the dimension of the gauge group is 14.

In order to determine the gauge group, recall that $[x,y,-]$ acts as a derivation:
\beq
[x,y,zw]=[x,y,z]w+z[x,y,w].
\eeq
This means that, if we define $\lambda: \scA\to \scA$ by $\lambda\cdot e^a=\lambda_{cd} f^{cdb}{}_a e^b$, we have 
\beq
\lambda(z w)=(\lambda z) w+z (\lambda w)
\eeq
This means that $\lambda$ is a derivation. Namely, gauge transformations are contained in the set of derivations of $\bO$. Now it is known since long ago \cite{Cartan} that derivation of $\bO$ is nothing but the exceptional Lie algebra $G_2$:
\beq
\mathfrak{der}(\bO)=G_2 \subset SO(7) \subset SO(8).
\eeq
This means that the gauge group of our theory is $G_2$, as claimed above. \\

We have constructed a three-dimensional theory with $\scN=2$ supersymmetry and with $SU(4)\times U(1)$ global symmetry. What is the physical meaning of this fact? First, this is inconsistent (at least in our example) with statement in \cite{CS} that we have $SU(4) \times U(1)$ R-symmetry. Indeed, the classification of $\scN=6$ theories in \cite{ST} tells us that theories with $\scN=6$ should have gauge group $SU(n)\times U(1), Sp(n)\times U(1), SU(n)\times SU(n)$ and $SU(n)\times SU(m)\times U(1)$ with possibly additional $U(1)$'s, and no $G_2$ gauge groups are allowed. Thus $SU(4)\times U(1)$ should be considered as a global symmetry, rather than a R-symmetry.

Second, the appearance of nonassociative algebras and generalized Lie 3-algebras strongly suggest the connection to membrane physics. Unfortunately, the connection of generalized Lie 3-algebras and worldvolume theories on M2-branes are currently not known, but our theory is certainly the possible candidate for the worldvolume of M2-branes with gauge group $G_2$. 
If this is indeed the case, this would be a novel way to realize exceptional gauge symmetry in M-theory \footnote{Another way to realize exceptional gauge symmetries in M-theory is to consider compactification of M-theory on K3 surfaces with ADE singularities.}.

\section{Summary and discussions}\label{sec.conc}

In this paper, we gave an explicit example of a generalized Lie 3-algebra, using one of the most famous example of non-associative algebras, namely the octonions. When combined with the result of \cite{CS}, we have three-dimensional Chern-Simons-matter theories with gauge group $G_2$ and with global symmetry $SU(4)\times U(1)$. The appearance of generalized Lie 3-algebra suggests that this theory is a possible candidate for the theory on M2-branes with exceptional gauge group $G_2$.

This raises many question which needs further exploration. First, it would be interesting to study the moduli space of this theory. Just as in the BLG/ABJM theories and their variants \cite{ABJM,moduli}, this will give invaluable information about the physical interpretation of our theory. Related to this question is the reduction to type IIA theory (probably along the lines of \cite{M2toD2}). The $G_2$ gauge symmetry should be broken into a smaller symmetry group in this process.

We can also envisage possible generalizations. In this paper, 
we have taken octonions as an example, but our strategy should be much more general. As written in Appendix, much of our construction applies to a wider class of non-associative algebras known as alternative algebras, and it would be interesting to consider
generalizations to other alternative algebras. Aside from alternative algebras, another interesting class of nonassociative algebras is
 Jordan algebras. Interestingly, other exceptional gauge groups $F_4, E_6 , E_7$ and $E_8$ are related to
Jordan algebras in interesting way \cite{Baez} (for example, $F_4$ is equal to Isom$(\bO\bP^2)$, where $\bO\bP^2$ is an octonionic projective plane), we have the hope of constructing theories with these exceptional gauge groups.

Finally, in BLG theory, we have obtained a class of theories by relaxing the condition
of positivity of the metric \cite{Lorentzian}. The same strategy should also work of our case as well.
For example, we have an algebra split-octonion, which has signature (4, 4), as contrast to
the ordinary octonions whose signature is (8, 0). Although the issue of unitarity is subtle,
these algebras might has some role to play in physics.

\section*{Note Added}
After the completion of version 1 of this paper, we received a paper \cite{recent} which discusses general theory of generalized Lie 3-algebras. In particular, they show that there exists a one-to-one correspondence between a generalized metric Lie 3-algebra and a pair consisting of a Lie algebra and its faithful orthogonal representation. In this general framework, our example is constructed from a Lie algebra $G_2$ and the octonions as its eight-dimensional representation, which in turn decomposes as $\bf{8}=\bf{7}+\bf{1}$ as can be seen in \eqref{eq:decouple}. Still, it still seems highly non-trivial to check explicitly that the 3-brackets defined from equation (9) of \cite{recent} matches with our definitions of 3-brackets. We thank Jos$\acute{\mathrm{e}}$~Figueroa-O'Farrill for valuable comments on this point.

\section*{Acknowledgments}

We thank Sergey Cherkis, Jos$\acute{\mathrm{e}}$ Figueroa-O'Farrill, Hiroyuki Fuji, Yosuke Imamura, Neil Lambert, Hirosi Ooguri, Christian S\"{a}mann, Yuji Tachikawa, Seiji Terashima and Futoshi Yagi for valuable comments, discussions and correspondence. This research is
supported by JSPS fellowships for Young Scientists.

\appendix

\section{Alternative Algebras}\label{sec.appendix}
In this section we collect some useful facts about alternative algebras, some of which are used in the main text. See \cite{schafer} for details of these algebras.

%
An alternative algebra is a special class of non-associative algebra $\scA$ which satisfies 
\beq
(x^2) y=x(xy), ~ y(x^2)=(yx)x
\eeq
for all $x,y\in \scA$. These two equations are known respectively as the left and right alternative laws.
From the multiplication table of octonions shown in Table \ref{tbl.oct}, it is easy to show that the octonions satisfy these conditions.

If you define an associator $\langle x,y,z \rangle$ by
\beq
\langle x,y,z \rangle = x(yz)-(xy)z,
\eeq
then this is equivalent to
\beq
\langle x,x,y \rangle=\langle y,x,x \rangle=0.
\eeq
From these identities, we have
\beq
0=\langle x+y,x+y,z \rangle =\langle x,y,z \rangle+\langle y,x,z \rangle,
\eeq
and thus $\langle x,y,z \rangle=-\langle y,x,z \rangle$. Proceeding in a similar way we can prove that associator is antisymmetric:
\beq
\langle z_{\sigma(1)},x_{\sigma(2)},x_{\sigma(3)} \rangle=\mbox{sgn}(\sigma) \langle x_1,x_2,x_3 \rangle. \label{eq.alt}
\eeq
In particular means that we have
\beq
\langle x,y,x \rangle=0,
\eeq
or
\beq
(xy)x=x(yx),
\eeq
which is called the flexible identity.





\begin{thebibliography}{99}
\parskip=-3pt

\bibitem{CS}
  S.~Cherkis and C.~Saemann,
  ``Multiple M2-branes and Generalized 3-Lie algebras,''
  arXiv:0807.0808 [hep-th].

\bibitem{BLG}
  J.~Bagger and N.~Lambert,
  Phys.\ Rev.\  D {\bf 75}, 045020 (2007)
  [arXiv:hep-th/0611108];
％
  Phys.\ Rev.\  D {\bf 77}, 065008 (2008)
  [arXiv:0711.0955 [hep-th]];
%
  JHEP {\bf 0802}, 105 (2008)
  [arXiv:0712.3738 [hep-th]].

  A.~Gustavsson,
  ``Algebraic structures on parallel M2-branes,''
  arXiv:0709.1260 [hep-th];

\bibitem{no-go}
  G.~Papadopoulos,
  ``M2-branes, 3-Lie Algebras and Plucker relations,''
  arXiv:0804.2662 [hep-th];

  J.~P.~Gauntlett and J.~B.~Gutowski,
  ``Constraining Maximally Supersymmetric Membrane Actions,''
  arXiv:0804.3078 [hep-th].

\bibitem{Lorentzian}
  J.~Gomis, G.~Milanesi and J.~G.~Russo,
  ``Bagger-Lambert Theory for General Lie Algebras,''
  arXiv:0805.1012 [hep-th].

  S.~Benvenuti, D.~Rodriguez-Gomez, E.~Tonni and H.~Verlinde,
  ``N=8 superconformal gauge theories and M2 branes,''
  arXiv:0805.1087 [hep-th].

  P.~M.~Ho, Y.~Imamura and Y.~Matsuo,
  ``M2 to D2 revisited,''
  arXiv:0805.1202 [hep-th].

\bibitem{unitary}
  M.~A.~Bandres, A.~E.~Lipstein and J.~H.~Schwarz,
  ``Ghost-Free Superconformal Action for Multiple M2-Branes,''
  arXiv:0806.0054 [hep-th].

  J.~Gomis, D.~Rodriguez-Gomez, M.~Van Raamsdonk and H.~Verlinde,
  ``The Superconformal Gauge Theory on M2-Branes,''
  arXiv:0806.0738 [hep-th].

\bibitem{D2toD2}
  B.~Ezhuthachan, S.~Mukhi and C.~Papageorgakis,
  ``D2 to D2,''
  arXiv:0806.1639 [hep-th].

\bibitem{ABJM}
  O.~Aharony, O.~Bergman, D.~L.~Jafferis and J.~Maldacena,
  ``N=6 superconformal Chern-Simons-matter theories, M2-branes and their
  gravity duals,''
  arXiv:0806.1218 [hep-th].

\bibitem{BL4}
  J.~Bagger and N.~Lambert,
  ``Three-Algebras and N=6 Chern-Simons Gauge Theories,''
  arXiv:0807.0163 [hep-th].

\bibitem{recent}
  P.~de Medeiros, J.~Figueroa-O'Farrill, E.~Mendez-Escobar and P.~Ritter,
  ``On the Lie-algebraic origin of metric 3-algebras,''
  arXiv:0809.1086 [hep-th].

\bibitem{ST}
  M.~Schnabl and Y.~Tachikawa,
  ``Classification of N=6 superconformal theories of ABJM type,''
  arXiv:0807.1102 [hep-th].

\bibitem{Hosomichi}
  K.~Hosomichi, K.~M.~Lee, S.~Lee, S.~Lee and J.~Park,
  ``N=5,6 Superconformal Chern-Simons Theories and M2-branes on Orbifolds,''
  arXiv:0806.4977 [hep-th].

\bibitem{Bergshoeff}
  E.~A.~Bergshoeff, O.~Hohm, D.~Roest, H.~Samtleben and E.~Sezgin,
  ``The Superconformal Gaugings in Three Dimensions,''
  arXiv:0807.2841 [hep-th].

\bibitem{Gustavsson}  
A.~Gustavsson,
  ``Selfdual strings and loop space Nahm equations,''
  arXiv:0802.3456 [hep-th].

\bibitem{oldoct}
  F.~Englert,
  Phys.\ Lett.\  B {\bf 119}, 339 (1982).

  B.~de Wit and H.~Nicolai,
  Nucl.\ Phys.\  B {\bf 231}, 506 (1984).

  M.~J.~Duff, B.~E.~W.~Nilsson and C.~N.~Pope,
  Phys.\ Rept.\  {\bf 130}, 1 (1986).

  M.~J.~Duff, J.~M.~Evans, R.~R.~Khuri, J.~X.~Lu and R.~Minasian,
  Phys.\ Lett.\  B {\bf 412}, 281 (1997)
  [Nucl.\ Phys.\ Proc.\ Suppl.\  {\bf 68}, 295 (1998)]
  [arXiv:hep-th/9706124].

  E.~G.~Floratos and G.~K.~Leontaris,
  Nucl.\ Phys.\  B {\bf 512}, 445 (1998)
  [arXiv:hep-th/9710064].

\bibitem{Mauri}
  A.~Morozov,
  JHEP {\bf 0805}, 076 (2008)
  [arXiv:0804.0913 [hep-th]].


  A.~Mauri and A.~C.~Petkou,
  ``An N=1 Superfield Action for M2 branes,''
  arXiv:0806.2270 [hep-th].

  M.~Axenides and E.~Floratos,
  ``Nambu-Lie 3-Algebras on Fuzzy 3-Manifolds,''
  arXiv:0809.3493 [hep-th].




\bibitem{schafer}
R.~D.~ Schafer, ``An Introduction to Nonassociative Algebras,'' Dover, 1995

\bibitem{Nambu}
  Y.~Nambu,
  Phys.\ Rev.\  D {\bf 7}, 2405 (1973).

\bibitem{Nilsson}
  B.~E.~W.~Nilsson and J.~Palmkvist,
  ``Superconformal M2-branes and generalized Jordan triple systems,''
  arXiv:0807.5134 [hep-th].

\bibitem{Cartan}
E.~Cartan, 
Ann. Sci. Ecole Norm. Sup. {\bf 31} (1914), 255-262

\bibitem{moduli}
  N.~Lambert and D.~Tong,
  ``Membranes on an Orbifold,''
  arXiv:0804.1114 [hep-th].

J. Distler, S. Mukhi, C. Papageorgakis, M. Van Raamsdonk,
``M2-branes on M-folds,''
arXiv:0804.1256 [hep-th].

H.~Fuji, S.~Terashima and M.~Yamazaki,
  ``A New N=4 Membrane Action via Orbifold,''
  arXiv:0805.1997 [hep-th].

  K.~Hosomichi, K.~M.~Lee, S.~Lee, S.~Lee and J.~Park,
  Multiplets,''
  JHEP {\bf 0807}, 091 (2008)
  [arXiv:0805.3662 [hep-th]].

  P.~De Medeiros, J.~M.~Figueroa-O'Farrill and E.~Mendez-Escobar,
  JHEP {\bf 0807}, 111 (2008)
  [arXiv:0805.4363 [hep-th]].


  M.~Benna, I.~Klebanov, T.~Klose and M.~Smedback,
  ``Superconformal Chern-Simons Theories and AdS$_4$/CFT$_3$ Correspondence,''
  JHEP {\bf 0809}, 072 (2008)
  [arXiv:0806.1519 [hep-th]].




  Y.~Imamura and K.~Kimura,
  ``On the moduli space of elliptic Maxwell-Chern-Simons theories,''
  arXiv:0806.3727 [hep-th].

  K.~Hosomichi, K.~M.~Lee, S.~Lee, S.~Lee and J.~Park,
  JHEP {\bf 0809}, 002 (2008)
  [arXiv:0806.4977 [hep-th]].


  S.~Terashima and F.~Yagi,
  ``Orbifolding the Membrane Action,''
  arXiv:0807.0368 [hep-th].

  D.~Martelli and J.~Sparks,
  ``Moduli spaces of Chern-Simons quiver gauge theories,''
  arXiv:0808.0912 [hep-th].

  A.~Hanany and A.~Zaffaroni,
  ``Tilings, Chern-Simons Theories and M2 Branes,''
  arXiv:0808.1244 [hep-th].


  K.~Ueda and M.~Yamazaki,
  ``Toric Calabi-Yau four-folds dual to Chern-Simons-matter theories,''
  arXiv:0808.3768 [hep-th].


  Y.~Imamura and K.~Kimura,
  ``Quiver Chern-Simons theories and crystals,''
  arXiv:0808.4155 [hep-th].


  A.~Hanany, D.~Vegh and A.~Zaffaroni,
  ``Brane Tilings and M2 Branes,''
  arXiv:0809.1440 [hep-th].

\bibitem{M2toD2}
  S.~Mukhi and C.~Papageorgakis,
  ``M2 to D2,''
  arXiv:0803.3218 [hep-th];

\bibitem{Baez}
J.~Baez, 
Bull. Amer. Math. Soc. {\bf 39} (2002), 145-205 [arXiv:math.RA/0105155]; Errata {\it ibid.} {\bf 42} (2005), 213
 .
\end{thebibliography}
\end{document}